\documentclass[showpacs,aps,prd,nofootinbib,floatfix,amsmath,amssymb,twocolumn]{revtex4}
\usepackage{mathrsfs}
\usepackage{graphicx}
\usepackage[dvipsnames]{xcolor}
\usepackage{dsfont}
\begin{document}

\makeatletter
%Feynman slash
\newbox\slashbox \setbox\slashbox=\hbox{$/$}
\newbox\Slashbox \setbox\Slashbox=\hbox{\large$/$}
\def\pFMslash#1{\setbox\@tempboxa=\hbox{$#1$}
  \@tempdima=0.5\wd\slashbox \advance\@tempdima 0.5\wd\@tempboxa
  \copy\slashbox \kern-\@tempdima \box\@tempboxa}
\def\pFMSlash#1{\setbox\@tempboxa=\hbox{$#1$}
  \@tempdima=0.5\wd\Slashbox \advance\@tempdima 0.5\wd\@tempboxa
  \copy\Slashbox \kern-\@tempdima \box\@tempboxa}
\def\FMslash{\protect\pFMslash}
\def\FMSlash{\protect\pFMSlash}
\def\miss#1{\ifmmode{/\mkern-11mu #1}\else{${/\mkern-11mu #1}$}\fi}
%%%% Uso:  \pFMSlash{p}
\makeatother

%\tightenlines
\title{New physics in $WW\gamma$ at one loop via Majorana neutrinos}

\author{Eduardo Mart\'inez$^{(a)}$}\author{Javier Monta\~no-Dom\'inguez$^{(b)}$}\author{H\'ector Novales-S\'anchez$^{(a)}$}
\author{M\'onica Salinas$^{(c)}$}
\affiliation{
$^{(a)}$Facultad de Ciencias F\'isico Matem\'aticas, Benem\'erita Universidad Aut\'onoma de Puebla, Apartado Postal 1152 Puebla, Puebla, M\'exico\\$^{(b)}$C\'atedras Conacyt-Facultad de Ciencias F\'isico Matem\'aticas,
Universidad Michoacana de San Nicol\'as de Hidalgo,
Av. Francisco J. M\'ugica s/n, C. P. 58060, Morelia, Michoac\'an, M\'exico.\\$^{(c)}$Departamento de F\'isica, Centro de Investigaci\'on y de Estudios Avanzados del IPN,
Apartado Postal 14-740,07000 Ciudad de M\'exico, M\'exico}

\begin{abstract}
Current experimental data guarantees the presence of physics beyond the Standard Model in the neutrino sector. The responsible physical description might show itself through virtual effects on low-energy observables. In particular, massive neutrinos are able to produce contributions to the triple gauge coupling $WW\gamma$. The present paper deals with the calculation, estimation and analysis of one-loop contributions from Majorana neutrinos to the Lorentz-covariant $WW\gamma$ parametrization. Our calculations show that CP-odd effects vanish exactly, whereas CP-even contributions, $\Delta\kappa$ and $\Delta Q$, remain. According to our estimations, the effects from heavy neutrinos with masses in the range of hundreds of GeVs dominate over those from light neutrinos. This investigation shows that contributions from heavy Majorana neutrinos to the anomaly $\Delta\kappa$ could be as large as $\sim{\cal O}(10^{-3})$, one order of magnitude below the Standard-Model contribution. We find that the International Linear Collider, sensitive to triple gauge couplings participating in $WW$ production, might measure these effects in electron-positron collisions at a center-of-mass energy of $800\,{\rm GeV}$, as long as heavy-neutrino masses are $\gtrsim300\,{\rm GeV}$ and below $\sim1500\,{\rm GeV}$. 
\end{abstract}


\maketitle

\section{Introduction}
Nowadays, the Standard Model~\cite{SMGlashow,SMSalam,SMWeinberg} (SM) remains our best description of fundamental physics. While most experimental data leans towards this field-theory formulation, phenomena beyond the Standard Model (BSM) have been confirmed. Experimental data has established the presence of new physics within the neutrino sector, namely, as opposed to the SM, neutrinos are massive and mix with each other. Nearly 70 years after the introduction of neutrinos by Pauli, the first evidence of neutrino mass materialized though the measurement of neutrino oscillations~\cite{Pontecorvo}, by the Kamiokande Collaboration~\cite{KamiokandeNobel}, which was confirmed four years later by the Sudbury Neutrino Observatory~\cite{SNONobel}. Before that, the lack of experimental evidence in favor of neutrino mass and the observed absence of right-handed neutrino states, in agreement with the Weyl description of massless fermions, motivated the assumption of massless neutrinos in the definition of the SM. The determination that neutrinos are massive bears great relevance {\it per se}, and yet it comes along with a quite interesting physical consequence: the possibility that the correct characterization of neutrinos is not provided by Dirac spinors~\cite{DiracNobel}, but by Majorana fields~\cite{Majorana} instead, in which case neutrino fields, $\nu$, and their charge-conjugate fields, $\nu^{\rm c}$, coincide with each other. This is an open problem, whose answer may come from the elusive neutrinoless double beta decay, which, if observed, would incarnate strong evidence upholding Majorana neutrinos, as this physical process is forbidden if such particles are Dirac-like. So far, there is no experimental support implying the occurrence of the neutrinoless double beta decay, despite efforts by several experimental collaborations~\cite{nodoublebeta,CUPIDMo0nu2beta,CUORE0nu2beta,GERDA0nu2beta,MAJORANA0nu2beta,EXO0nu2beta,KamLANDZen0nu2beta}. \\

Contrary to the case of the Dirac field, which provides a proper description of SM fermions, Majorana mass terms do not preserve lepton number, but they rather violate it in two units. Such a framework allows for the occurrence of the Weinberg operator\footnote{It has been pointed out that lepton-number conserving generalizations of the Weinberg operator can be set, but they require the introduction of extra scalar fields~\cite{CSV1,CSV2}.}~\cite{Weinbergoperator}, an effective-Lagrangian term of mass dimension 5, built solely upon SM fields and compatible with a universe in which the seesaw mechanism is responsible for the generation and definition of neutrino masses~\cite{MoSe1,MoSe2}. If nature turns out to be such that known neutrinos get their masses by a seesaw mechanism, heavy neutrinos must exist, with masses proportional to some high-energy scale associated to spontaneous symmetry breaking. The theoretical framework for the present investigation has been set in Ref.~\cite{Pilaftsis}, where the SM is extended, to a certain level of generality, by the inclusion of Dirac mass terms, generated as a consequence of the SM electroweak symmetry breaking~\cite{EnBr,Higgs}, and Majorana mass terms, presumably originating from a fundamental description of nature in which spontaneous symmetry breaking at some high-energy scale, $w$, takes place. In this context, the necessary conditions are met for a type-1 seesaw mechanism to operate, which defines neutrino masses at the tree level, with the distinctive seesaw-mass profile. Light-neutrino masses, being as tiny as $m_\nu<0.8\,{\rm eV}$~\cite{KATRINnumass}, require a large suppression provided by the energy scale $w$, characterizing the fundamental formulation. Such a large energy scale greatly increases the values of heavy-neutrino masses and pushes their effects out from experimental sensitivity. The author of Ref.~\cite{Pilaftsis} eluded this issue by imposing a condition aimed at the elimination of tree-level masses of light neutrinos. These masses were generated through radiative corrections instead, thus enabling heavy-neutrino masses to be smaller and, consequently, improving expectations regarding the size of the corresponding new-physics effects.\\

BSM effects from virtual heavy neutrinos on SM observables might be within the reach of current experimental sensitivity, which motivates the present work. We aim at the calculation, estimation, and analysis of the contributions from massive Majorana neutrinos, both light and heavy, to the $WW\gamma$ vertex at one loop, in the framework given by the neutrino model of Ref.~\cite{Pilaftsis}. The general parametrization of such an interaction was given long ago~\cite{BGL}, from which $W$-boson production mechanisms in linear colliders~\cite{HPZH,BaZe} and from $pp$ collisions~\cite{PaPh} were explored, in the context of the electroweak SM. Since then, SM extensions have been considered in order to estimate their one-loop contributions to $WW\gamma$. We calculate new-physics contributions by assuming both external $W$ bosons to be on shell, but keeping the external photon off the mass shell. From the form factors constituting the $WW\gamma$ parametrization, which yield the definitions of the electromagnetic properties of the $W$ boson, both CP-even and CP-odd, we identify the corresponding contributions engendered by the aforementioned neutrino model. Nonetheless, an exact cancellation of CP-odd contributions happens, thus leaving CP-even effects only. Moreover, the remaining contributions are found to be ultraviolet finite and gauge invariant. Pointing towards an estimation of the impact of the new physics, we consider nearly-degenerate heavy-neutrino mass spectra, which is required by this Majorana-neutrino model~\cite{Pilaftsis}. Our calculations and estimations show that effects from virtual heavy neutrinos are dominant over contributions from their light counterparts, by about 1 order of magnitude, for a wide range of heavy-neutrino masses. In optimistic scenarios, in which heavy-neutrino masses lie between $\sim300\,{\rm GeV}$ and $\sim1500\,{\rm GeV}$, the effects from virtual heavy neutrinos are found to be smaller than the one-loop SM contributions by $\sim1$ order of magnitude. Furthermore, our analysis suggest that new-physics contributions might be within the reach of the sensitivity of the International Linear Collider, expected to set bounds as stringent as ${\cal O}(10^{-4})$ on the CP-even anomalous couplings of the vertex $WW\gamma$, from the process $e^+e^-\to W^+W^-$, at a center-of-mass energy (CME) of $800\,{\rm GeV}$. \\

This paper has been organized as follows: we establish our framework by discussing the neutrino model of Ref.~\cite{Pilaftsis} in Section~\ref{theory}, where all the couplings and approximations, relevant for our calculation, are covered; then, in Section~\ref{calculation}, the parametrization of $WW\gamma$ is discussed, which is then followed by the calculation of the analytic expressions for the new-physics contributions to the form factors of such a coupling; the previously-obtained results are then utilized in order to perform numerical estimations in Section~\ref{analysis}, which are analyzed and discussed; finally, a summary of the paper and our conclusions are presented in Section~\ref{sumandconc}. \\

\section{Majorana neutrinos and the SM}
\label{theory}
Neutrinos are the only known existing elementary fermions characterized by electric neutrality\footnote{Note, however, that Ref.~\cite{StTo} has explored the possibility of millicharged neutrinos.}. Since electric charge is the mean to distinguish particles from their corresponding antiparticles, massive neutrinos meet the conditions to be depicted by Majorana spinors~\cite{Majorana}, which fulfill the so-called Majorana condition, $\nu^{\rm c}=\nu$. Even though the experimentally-confirmed phenomenon of neutrino oscillations, first proposed by Pontecorvo~\cite{Pontecorvo}, requires neutrinos to be massive, with all neutrino masses different of each other, it provides no information regarding whether these particles fit the Dirac o the Majorana description. On the other hand, the electromagnetic properties of neutrinos have been shown to differ in these cases~\cite{BGS}. Nowadays, the sensitivity of experiments aimed at the measurement of electromagnetic properties of neutrinos lies far away from the prediction of the SM minimally extended by massive neutrinos~\cite{GiKi}, by about 8 orders of magnitude~\cite{BGS}, thus meaning that BSM physics is the only hope to measure such electromagnetic interactions in the near future. The conventional probe of the Majorana nature of neutrinos is the neutrinoless double beta decay, which can occur only as long as neutrinos are Majorana fermions, but is not allowed if these particles are Dirac like. If neutrinos are Majorana, this physical process must be quite rare in nature, for no signals have been found in experiments so far, even though several collaborations have been pursuing it for decades~\cite{nodoublebeta,CUPIDMo0nu2beta,CUORE0nu2beta,GERDA0nu2beta,MAJORANA0nu2beta,EXO0nu2beta,KamLANDZen0nu2beta,PDG}. The GERDA Collaboration and the KamLAND-Zen Collaboration have established lower bounds of order $10^{26}$yr on the neutrinoless double-beta decay half-life~\cite{GERDA0nu2beta,KamLANDZen0nu2beta}.
\\

Consider a BSM model, characterized by some Lagrangian density ${\cal L}_{\rm BSM}$, upon which two phases of spontaneous symmetry breaking operate, namely, a first stage of symmetry breaking at some high-energy scale, here denoted by $w$, and, then, the Brout-Englert-Higgs mechanism~\cite{EnBr,Higgs}, through which the electroweak-SM gauge group, ${\rm SU}(2)_L\otimes{\rm U}(1)_Y$, is broken down into the electromagnetic group, ${\rm U}(1)_e$, at $v=246\,{\rm GeV}$. Instances of field-theory formulations in which this happens are left-right symmetric models~\cite{PaSa,MoPa1,MoPa2}, based on the gauge group ${\rm SU}(2)_L\otimes{\rm SU}(2)_R\otimes{\rm U}(1)_Y$, 331 models~\cite{PiPl,Frampton}, endowed with ${\rm SU}(3)_C\otimes{\rm SU}(3)_L\otimes{\rm U}(1)_X$ symmetry, and grand unification models~\cite{GeGl,GQW}, defined by larger gauge groups, such as ${\rm SU}(5)$ and ${\rm SO}(10)$. Then assume that this chain of events gives rise to the Lagrangian
\begin{equation}
{\cal L}_{\rm BSM}={\cal L}_{\rm mass}^\nu+{\cal L}^{\rm SM}_{\rm CC}+\cdots,
\label{LBSM}
\end{equation}
with the ellipsis standing for other terms, some of them specified by the BSM model under consideration. The Lagrangian term ${\cal L}_{\rm mass}^\nu$ is assumed to be given as
\begin{eqnarray}
{\cal L}_{\rm mass}^\nu=-\sum_{j=1}^3\sum_{k=1}^3\Big( \overline{\nu^0_{j,L}}(m_{\rm D})_{jk}\,\nu^0_{k,R}
\nonumber \\
+\frac{1}{2}\overline{\nu^{0\,{\rm c}}_{j,R}}(m_{\rm M})_{jk}\,\nu^0_{k,R} \Big)+{\rm H.c.},
\label{numassL}
\end{eqnarray}
where $\nu^0_{j,L}$ is a left-handed neutrino field and $\nu^0_{k,R}$ represents a neutrino field with right chirality, whereas $\psi^{\rm c}=C\overline{\psi}^{\rm T}$ is the charge-conjugate field of the spinor $\psi$, with $C$ the charge-conjugation matrix. Dirac mass-like terms, involving the $3\times3$ matrix $m_{\rm D}$ and its conjugate-transpose $m^\dag_{\rm D}$, are assumed to emerge from electroweak symmetry breaking. On the other hand, ${\cal L}^\nu_{\rm mass}$ involves Majorana mass-like terms as well, which are thought to emerge from spontaneous symmetry breaking at the high-energy scale $w$. The $3\times3$ matrix $m_{\rm M}$, lying within these terms, is symmetric, but general in any other regard. 
%\\ \\ \\ \\
By defining the $3\times1$ matrices
\begin{equation}
f_L=
\left(
\begin{array}{c}
\nu^0_{1,L}
\vspace{0.15cm}
\\
\nu^0_{2,L}
\vspace{0.15cm}
\\
\nu^0_{3,L}
\end{array}
\right),
\hspace{0.4cm}
F_L=
\left(
\begin{array}{c}
\nu^{0\,{\rm c}}_{1,R}
\vspace{0.15cm}
\\
\nu^{0\,{\rm c}}_{2,R}
\vspace{0.15cm}
\\
\nu^{0\,{\rm c}}_{3,R}
\end{array}
\right),
\end{equation}
and then denoting $f_R=C\overline{f_L}^{\rm T}$, $F_R=C\overline{F_L}^{\rm T}$, which emphasizes the chirality properties of these charge-conjugated fields, the neutrino-mass Lagrangian, ${\cal L}_{\rm mass}^\nu$, is written as
\begin{equation}
{\cal L}_{\rm mass}^\nu=-\frac{1}{2}(\overline{f_L}\hspace{0.2cm}\overline{F_L}){\cal M}\left(
\begin{array}{c}
f_R
\vspace{0.1cm}
\\
F_R
\end{array}
\right)+{\rm H.c.},
\end{equation}
where ${\cal M}$ is a $6\times6$ matrix, both complex and symmetric, given in terms of $3\times3$ matrix blocks as
\begin{equation}
{\cal M}=
\left(
\begin{array}{cc}
0 & m_{\rm D}
\\
m_{\rm D}^{\rm T} & m_{\rm M}
\end{array}
\right).
\end{equation}
In order to express the neutrino fields in the base of mass eigenspinors, a unitary diagonalization transformation is implemented through the unitary $6\times6$ matrix ${\cal U}_\nu$, given by $3\times3$ matrix blocks ${\cal U}_{kj}$ as
\begin{equation}
{\cal U}_\nu=
\left(
\begin{array}{cc}
{\cal U}_{11} & {\cal U}_{12}
\\
{\cal U}_{21} & {\cal U}_{22}
\end{array}
\right).
\label{unitaryU}
\end{equation}
The corresponding unitary transformation yields~\cite{Takagi}
\begin{equation}
{\cal U}^{\rm T}_\nu{\cal M}\,\,{\cal U}_\nu=\left(
\begin{array}{cc}
m_\nu & 0
\\
0 & m_N
\end{array}
\right),
\label{UMU}
\end{equation}
where $m_\nu$ and $m_N$ are $3\times3$ real-valued diagonal matrices, whose entries are $(m_\nu)_{jk}=m_{\nu_j}\delta_{jk}$ and $(m_N)_{jk}=m_{N_j}\delta_{jk}$, with $m_{\nu_j}>0$ and $m_{N_j}>0$. Thus, ${\cal L}_{\rm mass}^\nu$ acquires the form
\begin{equation}
{\cal L}_{\rm mass}^\nu=\sum_{j=1}^3\Big( -\frac{1}{2}m_{\nu_j}\overline{\nu_j}\nu_j-\frac{1}{2}m_{N_j}\overline{N_j}N_j \Big),
\label{Lnumass}
\end{equation}
comprising all the mass terms of the mass-eigenspinor neutrino fields $\nu_j$ and $N_j$, which satisfy $\nu_j^{\rm c}=\nu_j$, $N_j^{\rm c}=N_j$, meaning that they are Majorana-neutrino fields.
\\

The Lagrangian term ${\cal L}^{\rm SM}_{\rm CC}$, which is part of Eq.~(\ref{LBSM}), comprises the set of charged currents (CC) in which the SM $W$ boson participates. After implementing the change of basis defining the neutrino mass eigenspinors, ${\cal L}_{\rm CC}^{\rm SM}$ is expressed as
\begin{eqnarray}
{\cal L}_{\rm CC}^{\rm SM}=\sum_{\alpha}\sum_{j=1}^3\Big( \frac{g}{\sqrt{2}}{\cal B}_{\alpha\nu_j}W^-_\rho\,\overline{l_{\alpha}}\gamma^\rho P_L\nu_{j}
\nonumber \\
+\frac{g}{\sqrt{2}}{\cal B}_{\alpha N_j}W^-_\rho\,\overline{l_\alpha}\gamma^\rho P_LN_j\Big)+{\rm H.c.},
\label{CCs}
\end{eqnarray}
with the index $\alpha$ running over lepton flavors, that is $\alpha=e,\mu,\tau$. Moreover,
\begin{eqnarray}
{\cal B}_{\alpha\nu_j}=\sum_{k=1}^3V^\ell_{\alpha k}({\cal U}_{11}^*)_{kj},
\label{Balphanu}
\\
{\cal B}_{\alpha N_j}=\sum_{k=1}^3V^\ell_{\alpha k}({\cal U}_{12}^*)_{kj},
\end{eqnarray}
have been defined. In these equations, the $3\times3$ matrix $V^\ell$ is a lepton-sector analogue of the Kobayashi-Maskawa matrix~\cite{KM}, lying within the quark sector of the SM. In massive-neutrino BSM descriptions, such as those provided by the Weinberg operator~\cite{Weinbergoperator} and the minimally extended SM~\cite{GiKi}, lepton mixing is characterized by the matrix ${\cal U}_{\rm PMNS}$, dubbed the Pontecorvo-Maki-Nakagawa-Sakata (PMNS)~\cite{MNS,PontecorvoPMNS} matrix, which is unitary. Note that, by contrast, $V^\ell$ is not necessarily unitary. In models of Majorana neutrinos, ${\cal U}_{\rm PMNS}$ is determined by three mixing angles, $\theta_{12}$, $\theta_{13}$, and $\theta_{23}$, by the Dirac phase, $\delta_{\rm D}$, and by three Majorana phases, $\phi_1$, $\phi_2$, and $\phi_3$~\cite{GiKi}. The Majorana phase $\phi_1$ is usually taken $\phi_1=0$. The parametrization of the PMNS matrix for the case of Dirac neutrinos lacks the Majorana phases, so it is given by four parameters only~\cite{GiKi}. The set of coefficients ${\cal B}_{\alpha\nu_j}$ and ${\cal B}_{\alpha N_j}$ constitute the $3\times3$ matrices ${\cal B_\nu}$ and ${\cal B}_N$, respectively. Such matrices are conveniently arranged to define the $3\times6$ block matrix 
\begin{equation}
{\cal B}=\big( {\cal B}_\nu\hspace{0.15cm}{\cal B}_N \big),
\end{equation}
with entries
\begin{equation}
{\cal B}_{\alpha j}=
\left\{
\begin{array}{ll}
{\cal B}_{\alpha\nu_k} & {\rm if}\hspace{0.15cm}j=1,2,3,
\vspace{0.2cm}
\\
{\cal B}_{\alpha N_k} & {\rm if}\hspace{0.15cm}j=4,5,6,
\end{array}
\right.
\label{BvsB}
\end{equation}
where $\nu_k=\nu_1,\nu_2,\nu_3$ and $N_k=N_1,N_2,N_3$. The matrix ${\cal B}$ fulfills the conditions ${\cal B}{\cal B}^\dag={\bf 1}_3$ and ${\cal B}^\dag{\cal B}={\cal C}$, which are more explicitly expressed as
\begin{eqnarray}
\sum_{j=1}^6{\cal B}_{\alpha j}{\cal B}_{\beta j}^*=\delta_{\alpha\beta},
\\
\sum_{\alpha=e,\mu,\tau}{\cal B}_{\alpha j}^*{\cal B}_{\alpha k}={\cal C}_{jk}.
\end{eqnarray}
In these equations, $\delta_{\alpha\beta}=({\bf 1}_3)_{\alpha\beta}$, with ${\bf 1}_3$ the $3\times3$ identity matrix. Moreover,  the matrix ${\cal C}$, $6\times6$ sized, is given by its components as
\begin{equation}
{\cal C}_{jk}=\left\{
\begin{array}{ll}
{\cal C}_{\nu_i\nu_l}, {\rm if}\hspace{0.15cm}j=1,2,3\hspace{0.15cm}{\rm and}\hspace{0.15cm}k=1,2,3,
\vspace{0.2cm}
\\
{\cal C}_{\nu_iN_l}, {\rm if}\hspace{0.15cm}j=1,2,3\hspace{0.15cm}{\rm and}\hspace{0.15cm}k=4,5,6,
\vspace{0.2cm}
\\
{\cal C}_{N_i\nu_l}, {\rm if}\hspace{0.15cm}j=4,5,6\hspace{0.15cm}{\rm and}\hspace{0.15cm}k=1,2,3,
\vspace{0.2cm}
\\
{\cal C}_{N_iN_l}, {\rm if}\hspace{0.15cm}j=4,5,6\hspace{0.15cm}{\rm and}\hspace{0.15cm}k=4,5,6,
\end{array}
\right.
\end{equation}
with the definitions
\begin{eqnarray}
{\cal C}_{\nu_i\nu_l}=\sum_{j=1}^3({\cal U}_{11})_{ji}({\cal U}_{11}^*)_{jl}\equiv({\cal C}_{\nu\nu})_{il},
\label{Cnunu}
\\ \nonumber \\
{\cal C}_{\nu_iN_l}=\sum_{j=1}^3({\cal U}_{11})_{ji}({\cal U}_{12}^*)_{jl}\equiv({\cal C}_{\nu N})_{il},
\label{CnuN}
\\ \nonumber \\
{\cal C}_{N_i\nu_l}=\sum_{j=1}^3({\cal U}_{12})_{ji}({\cal U}_{11}^*)_{jl}\equiv({\cal C}_{N\nu})_{il},
\label{CNnu}
\\ \nonumber \\
{\cal C}_{N_iN_l}=\sum_{j=1}^3({\cal U}_{12})_{ji}({\cal U}_{12}^*)_{jl}\equiv({\cal C}_{NN})_{il},
\label{CNN}
\end{eqnarray}
in terms of the matrix blocks introduced in Eq.~(\ref{unitaryU}). Note that $\nu_j,\nu_k=\nu_1,\nu_2,\nu_3$ and $N_j,N_k=N_1,N_2,N_3$, so the matrix ${\cal C}$ is written in terms of $3\times3$ matrix blocks ${\cal C}_{\nu\nu}$, ${\cal C}_{\nu N}$, ${\cal C}_{N\nu}$, and ${\cal C}_{NN}$, which we defined in Eqs.~(\ref{Cnunu})-(\ref{CNN}), as
\begin{equation}
{\cal C}=\left(
\begin{array}{cc}
{\cal C}_{\nu\nu} & {\cal C}_{\nu N}
\vspace{0.2cm}
\\
{\cal C}_{N\nu} & {\cal C}_{NN}
\end{array}
\right).
\end{equation}
The property ${\cal C}{\cal C}^\dag={\cal C}$, expressed in terms of matrix components as $\sum_{i=1}^6{\cal C}_{ji}{\cal C}_{ki}^*={\cal C}_{jk}$, holds.
\\

Unitarity of the diagonalization matrix ${\cal U}_\nu$ and usage of Eq.~(\ref{UMU}) allows for the approximation~\cite{Pilaftsis}
%\begin{widetext}
\begin{equation}
{\cal U}_\nu\simeq
\left(
\begin{array}{cc}
{\bf 1}_3-\frac{1}{2}\xi^*\xi^{\rm T} & \xi^*\big( {\bf 1}_3-\frac{1}{2}\xi^{\rm T}\xi^* \big)J
\vspace{0.3cm}
\\
-\xi^{\rm T}\big( {\bf 1}_3-\frac{1}{2}\xi^*\xi^{\rm T} \big) & \big( {\bf 1}_3-\frac{1}{2}\xi^{\rm T}\xi^* \big)J
\end{array}
\right),
\label{approxU}
\end{equation}
%\end{widetext}
where the $3\times3$ matrix
\begin{equation}
\xi=m_{\rm D}m_{\rm M}^{-1},
\label{xidef}
\end{equation}
has been defined, with $|\xi_{jk}|<1$, in which case the smaller the entries of the $3\times3$ matrix $\xi$, the more reliable this approximation will be. Moreover, $J$ is a $3\times3$ matrix, both diagonal and unitary. From Eq.~(\ref{approxU}), we find that the matrix ${\cal B}$, defined by Eqs.~(\ref{Balphanu})-(\ref{BvsB}), is approximated as
\begin{equation}
{\cal B}\simeq
\textstyle
\Big( V^\ell\big( {\bf 1}_3-\frac{1}{2}\xi\,\xi^\dag \big)\hspace{0.3cm}V^\ell\,\xi\big( {\bf 1}_3-\frac{1}{2}\xi^\dag\xi \big)J^* \Big).
\end{equation}
If $m_{\rm D}\sim v$ and $m_{\rm M}\sim w$ are assumed, with $v\ll w$, type-1 seesaw mechanism takes place, defining light-neutrino masses $m_{\nu_j}\sim\frac{v^2}{w}$ and heavy-neutrino masses $m_{N_j}\sim w$, in line with Eq.~(\ref{UMU}). Masses of light neutrinos generated by the seesaw mechanism
%, as given by Eq.~(\ref{mnu}), 
place a severe lower bound on $w$,
%since for $M_{\rm D}M^{-1}_{\rm M}M_{\rm D}^{\rm T}\sim{\cal O}(1)$ this high-energy scale
which is restricted to be greater than $\sim10^{13}\,{\rm GeV}$, thus resulting in huge heavy-neutrino masses $m_{N_j}$. In the present paper, we rather follow Ref.~\cite{Pilaftsis}, in which the set of conditions $({\cal M}\,{\cal U}_\nu)_{jk}=0$, with $j=1,2,3,4,5,6$, are stated to be sufficient and necessary for the $k$-th neutrino mass to vanish at tree level, that is $m_{\nu_k}=0$ in Eq.~(\ref{UMU}).
%while keeping, on the other hand, nonzero seesaw masses of heavy neutrinos. 
On the other hand, the tree-level masses of heavy neutrinos are left untouched, so they are defined by the diagonalization of the mass matrix, Eq.~(\ref{UMU}), as~\cite{Pilaftsis}
\begin{equation}
m_N\simeq J\,m_{\rm M}\Big(1+\frac{1}{2}m^{-1}_{\rm M}\big( \xi^\dag m_{\rm D}+m_{\rm D}^{\rm T}\xi^* \big) \Big)J.
\label{mN}
\end{equation}
In this manner, very large values of heavy-neutrino masses and, consequently, quite suppressed contributions from new physics are avoided. Then, in Ref.~\cite{Pilaftsis}, massiveness of light neutrinos is achieved radiatively, via neutrino self-energy diagrams. Furthermore, that work provides and discusses the one-loop calculation of the light-neutrino mass matrix $m^{\rm loop}_\nu$. A noteworthy observation of such a calculation is that the smallness of light-neutrino masses comes about if the heavy neutrinos are nearly mass degenerate. 
\\

\section{One-loop effects from Majorana neutrinos on the vertex $WW\gamma$}
\label{calculation}
The present section is devoted to the analytic calculation, at the one-loop level, of the contributions from Feynman diagrams involving virtual Majorana neutrinos $\nu_j$ and $N_j$, emerged from the neutrino model discussed in Section~\ref{theory}, to the vertex $WW\gamma$. The presence of the CCs given in Eq.~(\ref{CCs}) yields vertices which make it possible for one-loop $WW\gamma$ diagrams, with such virtual neutrinos, to exist. The Lorentz-covariant structure of this vertex defines contributions to the electromagnetic moments of the $W$ boson, with a couple of them CP even and other two being CP odd. As we show and discuss below, contributions to CP-odd electromagnetic form factors vanish, thus leaving only CP-even effects.
\\

\subsection{Parametrization of $WW\gamma$}
We start our discussion by considering the effective Lagrangian~\cite{HPZH,BaZe}
\begin{equation}
{\cal L}^{WW\gamma}_{\rm eff}={\cal L}_{WW\gamma}^{\rm even}+{\cal L}^{\rm odd}_{WW\gamma},
\end{equation}
with the definitions
\begin{eqnarray}
&&
{\cal L}_{WW\gamma}^{\rm even}=-ie\,g_1\big(W^{+}_{\mu\nu}W^{-\mu}A^{\nu}-W_{\mu\nu}^{-}W^{+\mu}A^\nu\big)
\nonumber \\ &&
\hspace{0.6cm}
-ie\,\kappa\,W^+_\mu W^-_\nu F^{\mu\nu}-\frac{ie\lambda}{m_W^2}W^+_{\mu\nu}W^{-\nu}\hspace{0.000001cm}_{\rho}F^{\rho\mu},
\end{eqnarray}
\begin{eqnarray}
&&
{\cal L}_{WW\gamma}^{\rm odd}=-ie\,\tilde{\kappa}\,W_\mu^+W_\nu^-\tilde{F}^{\mu\nu}
\nonumber \\ &&
\hspace{1.5cm}
-\frac{ie\tilde{\lambda}}{m_W^2}W^+_{\mu\nu}W^{-\nu}\hspace{0.000001cm}_\rho\tilde{F}^{\rho\mu}
\end{eqnarray}
where $\kappa$, $\lambda$, $\tilde{\kappa}$, and $\tilde{\lambda}$ parametrize, at low energies, the effects of some new-physics formulation, whereas $e$ is the electric charge of a positron. While, as usual, $F_{\mu\nu}=\partial_\mu A_\nu-\partial_\nu A_\mu$ is the electromagnetic tensor, the dual tensor $\tilde{F}_{\mu\nu}=\frac{1}{2}\epsilon_{\mu\nu\rho\lambda}F^{\rho\lambda}$ is defined. The definitions $W^{\pm}_{\mu\nu}=\partial_\mu W^\pm_\nu-\partial_\nu W^\pm_\mu$ have been used as well. The criterion to separate ${\cal L}_{\rm eff}^{WW\gamma}$ into the lagrangian terms ${\cal L}^{\rm even}_{WW\gamma}$ and ${\cal L}^{\rm odd}_{WW\gamma}$ has been their properties with respect to the discrete transformation $CP$, under which, as indicated by our notation, such terms are either even or odd, respectively. Violation of CP symmetry holds great physical interest, for its presence is a requirement for the occurrence of baryon asymmetry in the universe, as stipulated by the Sakharov conditions~\cite{Sakharov}. Furthermore, since the only source of CP violation in the SM is the Kobayashi-Maskawa matrix, physical quantities not preserving CP symmetry incarnate means to look for BSM new physics and are quite relevant.
\\

\begin{figure}[ht]
\center
\includegraphics[width=6cm]{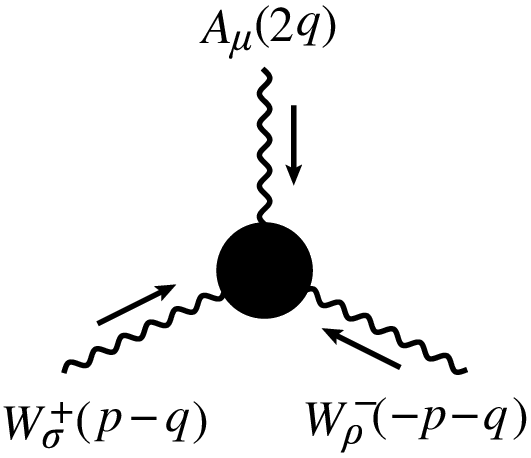}
\caption{\label{cnvWWgamma} Conventions for momenta in the $WW\gamma$ vertex.}
\end{figure}

With the objective of writing down, in momentum space, the expression for the vertex function $WW\gamma$ associated to the effective Lagrangian ${\cal L}_{\rm eff}^{WW\gamma}$, we provide Fig.~\ref{cnvWWgamma},
which displays the momenta conventions used for the $WW\gamma$ vertex throughout the present investigation. First introduced in Ref.~\cite{BGL}, this choice already takes into account momentum conservation of the whole vertex. With the assumptions that the external photon line is off shell and both $W$-boson external lines are on shell, the momenta conventions of Fig.~\ref{cnvWWgamma} are used to derive, from the lagrangian terms comprised by ${\cal L}_{WW\gamma}^{\rm even}$, the CP-even $WW\gamma$ vertex function
\begin{eqnarray}
&&
\Gamma^{\rm even}_{\sigma\rho\mu}=ie\Big( g_1\big( 2p_\mu\, g_{\sigma \rho}+4(q_\rho\,g_{\sigma\mu}-q_\sigma \,g_{\rho\mu}) \big)
\nonumber \\ &&
\hspace{1.5cm}
+\frac{4\,\Delta Q}{m_W^2}p_\mu\Big( q_\sigma q_\rho-\frac{q^2}{2}g_{\sigma\rho} \Big)
\nonumber \\ &&
\hspace{1.5cm}
+2\Delta\kappa\,(q_\rho\,g_{\sigma\mu}-q_\sigma\,g_{\rho\mu})
\Big),
\label{Geven}
\end{eqnarray}
with the definitions $\Delta\kappa=-g_1+\kappa+\lambda$ and $\Delta Q=-2\lambda$. Furthermore, ${\cal L}^{\rm odd}_{WW\gamma}$ leads to the CP-odd vertex function
\begin{eqnarray}
&&
\Gamma^{\rm odd}_{\sigma\rho\mu}=ie\Big( 2\Delta\tilde{\kappa}\,\epsilon_{\sigma\rho\mu\alpha}q^\alpha+\frac{4\Delta\tilde{Q}}{m_W^2}q_\rho\epsilon_{\sigma\mu\alpha\beta}p^\alpha q^\beta 
\nonumber \\ &&
\hspace{1cm}
+\tilde{g}_1\, p^\lambda\epsilon_{\sigma\rho\lambda\alpha}\big( q^2\delta^\alpha\hspace{0.000001cm}_\mu-q^\alpha q_\mu \big)
\Big),
\label{Godd}
\end{eqnarray}
where $\Delta\tilde{\kappa}=\tilde{\kappa}+\frac{m_W^2-2q^2}{m_W^2}\tilde{\lambda}$, $\Delta\tilde{Q}=-2\tilde{\lambda}$, and $\tilde{g}_1=-\frac{4\tilde{\lambda}}{m_W^2}$. Note that implementation of on-shell conditions on the photon external line eliminates contributions associated to the parameter $\tilde{g}_1$. At tree level in the SM, $g_1=1$, $\kappa=1$, and $\lambda=0$, whereas $\tilde{\kappa}=0$ and $\tilde{\lambda}=0$. These effective vertex functions fulfill Ward identities~\cite{Ward}, with respect to the electromagnetic field: $Q^\mu\Gamma^{\rm even}_{\sigma\rho\mu}=0$ and $Q^\mu\Gamma^{\rm odd}_{\sigma\rho\mu}=0$, where $Q=2q$ is the incoming momentum of the external photon. The electromagnetic properties of the $W$ boson comprise a set of four electromagnetic moments, which are defined, in terms of the parameters characterizing the $WW\gamma$ vertex functions of Eqs.~(\ref{Geven}) and (\ref{Godd}), as~\cite{HPZH,BaZe}
\begin{eqnarray}
\mu_W&=&\frac{e}{2m_W}(2+\Delta\kappa),
\\
Q_W&=&-\frac{e}{m^2_W}(1+\Delta\kappa+\Delta Q),
\\
\tilde{\mu}_W&=&\frac{e}{2m_W}\Delta\tilde{\kappa},
\\
\tilde{Q}_W&=&-\frac{e}{m_W^2}(\Delta\tilde{\kappa}+\Delta\tilde{Q}).
\end{eqnarray}
The $W$-boson electromagnetic moments $\mu_W$ and $Q_W$ are both even with respect to the $CP$ discrete transformation. These quantities are given the names ``magnetic dipole moment'' and ``electric quadrupole moment'', respectively. In the SM, these electromagnetic moments are nonzero at tree level. On the other hand, the ``electric dipole moment'', $\tilde{\mu}_W$, and the ``magnetic quadrupole moment'', $\tilde{Q}_W$, are CP odd and vanish at the tree level in the SM, so these quantities are interesting places to look for traces of BSM physics.
\\

\subsection{One-loop contributions to $WW\gamma$}
The contributions from the CCs lagrangian term ${\cal L}_{\rm CC}^{\rm SM}$, Eq.~(\ref{CCs}), to the $WW\gamma$ vertex function, $\Gamma^{WW\gamma}_{\sigma\rho\mu}$, are generated by the sum of Feynman diagrams
\begin{equation}
\Gamma^{WW\gamma}_{\sigma\rho\mu}=\sum_\alpha\sum_{k=1}^3
\Bigg(
\begin{gathered}
\vspace{-0.055cm}
\includegraphics[width=2.2cm]{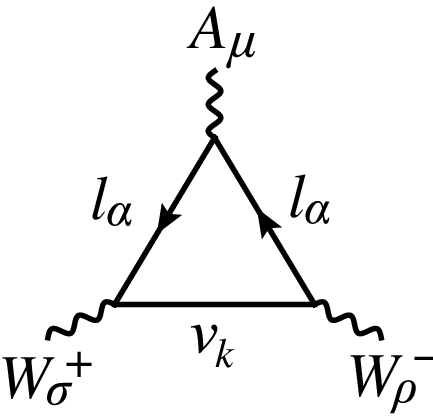}
\end{gathered}
\hspace{0.1cm}
+
\hspace{0.1cm}
\begin{gathered}
\vspace{-0.055cm}
\includegraphics[width=2.2cm]{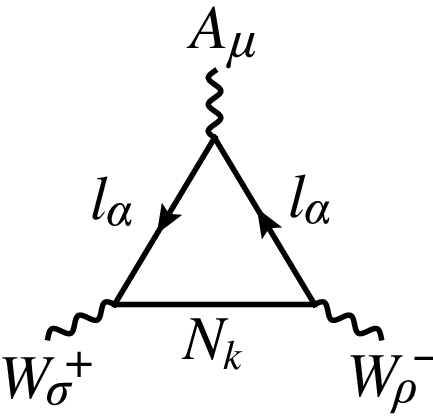}
\end{gathered}
\Bigg),
\label{vfsumofdiags}
\end{equation}
in accordance with the momenta conventions of Fig.~\ref{cnvWWgamma}. Also note that $\alpha=e,\mu,\tau$, so the sum $\sum_\alpha$ runs over the three lepton flavors. In general, calculations of amplitudes featuring neutrinos differ depending on whether, in the model at hand, such fermions are described by Dirac or Majorana fields. For starters, a main difference lies in the Feynman rules corresponding to each of these cases~\cite{DEHK}. Furthermore, lepton number violation allows for the occurrence of a larger number of Feynman diagrams if neutrinos are Majorana, as compared to the Dirac case. Let us point out that the present calculation, in which neutrinos are characterized by Majorana fields, does not entail any such additional diagrams. Moreover, those Feynman diagrams to be considered have the same structure as if neutrinos were Dirac type. Therefore, the calculation does not distinguish between the Dirac and Majorana cases, except for a possible role played by the Majorana phases. To perform the calculation, we have taken the external $W$ bosons on shell, whereas the external photon has been assumed to be off the mass shell. The set of contributing Feynman diagrams includes virtual effects from both light and heavy Majorana neutrinos, $\nu_j$ and $N_j$,  with Feynman rules determined from Eq.~(\ref{CCs}), in Section~\ref{theory}. Note that the superficial degree of divergence of any of the diagrams displayed in Eq.~(\ref{vfsumofdiags}) is 1, which warns us that ultraviolet divergences, growing as fast as linearly, might arise. With this in mind, we use the method of dimensional regularization~\cite{BoGi,tHVe} to regularize loop integrals, which, accordingly, are set in $D$ spacetime dimensions: $\int\frac{d^4k}{(2\pi)^4}\rightarrow\mu^{4-D}\int\frac{d^Dk}{(2\pi)^D}$, with $\mu$ the renormalization scale. To calculate the analytic expressions corresponding to these diagrams, we have followed the tensor-reduction method~\cite{PaVe}, implemented through the software Mathematica, by Wolfram, with the packages Feyncalc~\cite{Feyncalc1,Feyncalc2,Feyncalc3} and Package-X~\cite{PackageX}. Consequently, the analytic expressions of the contributions are given in terms of 1-point, 2-point, and 3-point Passarino-Veltman scalar functions.
\\

Let us rearrange the vertex function $\Gamma^{WW\gamma}_{\sigma\rho\mu}$, just given in Eq.~(\ref{vfsumofdiags}), as
\begin{equation}
\Gamma^{WW\gamma}_{\sigma\rho\mu}=\sum_\alpha\sum_{k=1}^6|{\cal B}_{\alpha k}|^2\Big( \Gamma^{\rm even}_{\sigma\rho\mu}(m_\alpha,m_k)+\Gamma^{\rm odd}_{\sigma\rho\mu}(m_\alpha,m_k) \Big).
\label{Gevenplusodd}
\end{equation}
In Eq.~(\ref{Gevenplusodd}), the contribution from the one-loop Feynman diagram involving the $\alpha$-th virtual charged lepton, with mass $m_\alpha$, and the $\nu_k$-th virtual neutrino has been splitted into CP-even and CP-odd contributions, denoted by $\Gamma^{\rm even}_{\sigma\rho\mu}(m_\alpha,m_k)$ and $\Gamma^{\rm odd}_{\sigma\rho\mu}(m_\alpha,m_k)$, respectively.
While our notation explicitly advises the reader about the dependence of the vertex function on the masses of charged leptons and neutrinos, let us point out that $\Gamma^{\rm even}_{\sigma\rho\mu}(m_\alpha,m_k)$ and $\Gamma^{\rm odd}_{\sigma\rho\mu}(m_\alpha,m_k)$ also comprise the $W$ boson mass, $m_W$, and the squared external-photon momentum, $Q^2$. In contrast with Eq.~(\ref{vfsumofdiags}), this expression of $\Gamma^{WW\gamma}_{\sigma\rho\mu}$ features the sum $\sum_{k=1}^6$, which takes all the neutrinos, both light and heavy, into account at once. For that reason, the masses of all the neutrinos have been generically denoted by $m_k$, where $k=1,2,3,4,5,6$, so $m_1=m_{\nu_1}$, $m_2=m_{\nu_2}$, and $m_3=m_{\nu_3}$ are the masses of light neutrinos, whereas the masses of the heavy neutrinos are $m_4=m_{N_1}$, $m_5=m_{N_2}$, and $m_6=m_{N_3}$. Similarly, factors ${\cal B}_{\alpha k}$, defined in Eq.~(\ref{BvsB}), have been utilized. 
\\

The structure of the CP-even vertex-function term $\Gamma_{\sigma\rho\mu}^{\rm even}(m_\alpha,m_k)$ matches Eq.~(\ref{Geven}), so it satisfies the corresponding Ward identity. This CP-even vertex function carries individual contributions $g_1^{\alpha k}$, $\Delta Q^{\alpha k}$, and $\Delta\kappa^{\alpha k}$. We found ultraviolet divergences nested within individual contributions $g^{\alpha k}_1$, which do not vanish from the total form factor $g_1=\sum_\alpha\sum_{k=1}^6|{\cal B}_{\alpha k}|^2g_1^{\alpha k}$. Nonetheless, this divergent contribution is absorbed by the SM through renormalization. The form-factor contributions $\Delta Q^{\alpha k}$ and $\Delta\kappa^{\alpha k}$, on the other hand, turned out to be free of ultraviolet divergences. We provide their explicit expressions in the Appendix. The full set of individual contributions $\Delta Q^{\alpha k}$ and $\Delta\kappa^{\alpha k}$ sum together to give the anomalous couplings
\begin{eqnarray}
\Delta\kappa&\displaystyle&=\sum_\alpha\sum_{k=1}^6|{\cal B}_{\alpha k}|^2\Delta\kappa^{\alpha k}.
\label{totDk}
\\
\Delta Q&\displaystyle&=\sum_\alpha\sum_{k=1}^6|{\cal B}_{\alpha k}|^2\Delta Q^{\alpha k},
\label{totDQ}
\end{eqnarray}
\\

As shown by Eq.~(\ref{Gevenplusodd}), the $WW\gamma$ vertex function $\Gamma^{WW\gamma}_{\sigma\rho\mu}$ includes CP-odd contributions as well, lying within the term $\Gamma^{\rm odd}_{\sigma\rho\mu}(m_\alpha,m_k)$. After processing via software tools, we got an expression with Lorentz-covariant structure
\begin{eqnarray}
&&\Gamma^{\rm odd}_{\sigma\rho\mu}(m_\alpha,m_k)=
\epsilon_{\mu\rho\sigma\alpha}p^\alpha F_1
\nonumber\\&&\hspace{1.7cm}
+p^\alpha q^\beta(q_\sigma\epsilon_{\mu\rho\alpha\beta} \,F_2
+q_\rho\epsilon_{\mu\sigma\alpha\beta} \,F_3),
\end{eqnarray}
where $F_1$, $F_2$, and $F_3$ are some functions of masses, given in terms of Passarino-Veltman scalar functions. Let us point out that the three form factors are ultraviolet finite. Merging Schouten identities~\cite{NeeVe} with the on-shell $W$-boson transversality conditions on momenta, we rewrite $\Gamma_{\sigma\rho\mu}^{\rm odd}(m_\alpha,m_k)$ as
\begin{eqnarray}
&&
\Gamma^{\rm odd}_{\sigma\rho\mu}(m_\alpha,m_k)=-\epsilon_{\sigma\rho\mu\alpha}\,p^\alpha F_1
\nonumber \\ && \hspace{2.2cm}
-q_\rho\epsilon_{\sigma\mu\alpha\beta}p^\alpha q^\beta(F_2+F_3)
\nonumber \\ &&\hspace{1.5cm}
-p^\lambda\epsilon_{\sigma\rho\lambda\alpha}(q^2\delta^\alpha\hspace{0.00001cm}_\mu-q^\alpha q_\mu)F_2.
%\nonumber\\ &&\hspace{2.5cm}
%-q_\rho\epsilon_{\sigma\mu\alpha\beta}p^\alpha q^\beta(F_2+F_3).
\label{cpoddquesalio}
\end{eqnarray}
When compared with Eq.~(\ref{Godd}), the second term of this expression, whose Lorentz-covariant structure is $q_\rho\epsilon_{\sigma\mu\alpha\beta}\,p^\alpha q^\beta$, leads to the identification $\Delta\tilde{Q}=\frac{i(F_2+F_3)m_W^2}{4e}$. We have verified that, even assuming the external photon to be off shell, the cancellation $F_2+F_3=0$ happens, so the CP-odd one-loop contribution $\Delta\tilde{Q}$ vanishes exactly. Moreover, Eq.~(\ref{cpoddquesalio}) shows no contribution to the form factor $\Delta\tilde{\kappa}$. On the other hand, we note that the off-shell contribution $\tilde{g}_1=\frac{iF_2}{e}$ is generated, in line with Eq.~(\ref{Godd}). The first term of Eq.~(\ref{cpoddquesalio}), characterized by the Lorentz-covariant structure $\epsilon_{\sigma\rho\mu\alpha}\,p^\alpha$, is not part of the general parametrization of the $WW\gamma$ vertex, shown in Eq.~(\ref{Godd}). Furthermore, this extra term, which does not vanish even when taking the vertex on the mass shell, spoils the Ward identity $Q^\mu\Gamma_{\sigma\rho\mu}^{\rm odd}(m_\alpha,m_k)=0$. Ward identities may fail if gauge symmetry, valid in some classical formulation, ceases to hold at the quantum level, in which case the theory is said to have anomalies~\cite{Adler,BeJa,Fujikawa}. Consistent field theories are anomaly free, as it is the case of the SM, where a set of conditions relating hypercharges are fulfilled in order to eliminate anomalies. Ward identities, customarily understood as a test of gauge invariance, often serve as a criterion to probe consistency of calculations of amplitudes, since failure of some Ward identity may hint at a mistake in the execution of a given calculation. However, there is another plausible reason for Ward identities not to occur in the context of an anomaly-free theory, namely, the presence of the chirality matrix, $\gamma_5$, which turns out to be incompatible with the method of dimensional regularization~\cite{tHVe,Jegerlenhner}. In order to tackle amplitudes involving the chirality matrix, modi operandi have been posed. For instance, the $\gamma_5$ is taken to anticommute with all the Dirac matrices $\gamma^\mu$, present in $D$ spacetime dimensions, in the so-called naive dimensional regularization. Another approach, by t'Hooft and Veltman~\cite{tHVe}, takes the $\gamma_5$ to anticommute with the four original Dirac matrices, $\gamma^0$, $\gamma^1$, $\gamma^2$, and  $\gamma^3$, while commutation relations with the rest of such matrices, in $D$ dimensions, are imposed to hold. Extensions of the t'Hooft-Veltman approach, such as the Breitenlohner-Maison scheme~\cite{BrMa}, are available as well. The author of Ref.~\cite{Jegerlenhner} claims that loop integrals involving the $\gamma_5$ can be regularized only as long as either the trace condition ${\rm tr}\{ \gamma^\mu\gamma^\nu\gamma^\rho\gamma^\sigma\gamma_5 \}\ne0$ or gauge invariance is abandoned. This author suggests, in practice, usage of the naive dimensional regularization, as this avoids both spurious anomalies and complications. Returning to our calculation, we have verified that the use of the naive dimensional regularization, which comes along with the trace condition ${\rm tr}\{ \gamma^\mu\gamma^\nu\gamma^\rho\gamma^\sigma\gamma_5 \}=0$, renders the CP-odd contribution zero, that is $\Gamma_{\sigma\rho\mu}^{\rm odd}(m_\alpha,m_k)=0$, in which case the Ward identity $Q^\mu\Gamma_{\sigma\rho\mu}^{\rm odd}(m_\alpha,m_k)=0$ is trivially fulfilled. On the other hand, after implementation of the cancellation $F_2+F_3=0$, commented above, Eq.~(\ref{cpoddquesalio}) suggests that traces ${\rm tr}\{ \gamma^\mu\gamma^\nu\gamma^\rho\gamma^\sigma\gamma_5 \}$ do not contribute to the electromagnetic form factors, so we can pragmatically ignore these traces by using the naive dimensional regularization. Moreover, both approaches coincide in that CP-odd electromagnetic form factors vanish, even if the photon is off shell. We arrive at the conclusion that this model of Majorana neutrinos does not produce any contribution to the CP-odd $WW\gamma$ anomalous couplings $\Delta\tilde{\kappa}$ and $\Delta\tilde{Q}$, so no contributions to the CP-odd electromagnetic moments of the $W$ boson, $\tilde{\mu}_W$ and $\tilde{Q}_W$, arise.\\

\section{Estimations and discussion}
\label{analysis}
The main purpose of the present section is the estimation of the one-loop contributions from the Majorana-neutrino model discussed in Section~\ref{theory} to the electromagnetic form factors characterizing the $WW\gamma$ vertex. As we previously showed, in Section~\ref{calculation}, only CP-even contributions are generated, which are given by the anomalous couplings $\Delta\kappa$ and $\Delta Q$, as displayed by Eq.~(\ref{Geven}). In Ref.~\cite{AKLPS}, a calculation and estimation of the one-loop contributions from the SM to the vertex $WW\gamma$, with the external photon taken off shell, was carried out. The authors of this work derived expressions in terms of the masses of the Higgs boson and the top quark, none of which had been measured at the time. They performed numerical estimations of $\Delta\kappa$ and $\Delta Q$, as functions on the external-photon quadratic momentum, from which they arrived at the conclusion that SM contributions to $\Delta\kappa$ could be as large as $10^{-2}$, meaning that any measurement of a larger value of this anomaly should be interpreted as a manifestation of BSM physics. Moreover, SM contributions to $\Delta Q$ were determined to be as large as $10^{-4}$. Shortly later, the SM effects on $WW\gamma$ were reexamined in Ref.~\cite{PaPhWWA}, in which, by usage of the so-called Pinch Technique~\cite{Cornwall,CoPa,PapavassiliouPT}, contributions to $\Delta\kappa$ and $\Delta Q$ were found to be gauge independent, ultraviolet and infrared finite, and well behaved for large off-shell photon momentum. The effects of BSM physics on the vertex $WW\gamma$ have been addressed as well. In Refs.~\cite{MTTR,RTT}, for instance, the 331 model was the framework to execute one-loop calculations of $WW\gamma$, whereas Ref.~\cite{LMMNTT} considered contributions to this vertex, at one loop, from a model of universal extra dimensions~\cite{ACD}. In Ref.~\cite{MRTT}, the one-loop contributions to the CP-odd electromagnetic moments $\tilde{\mu}_W$ and $\tilde{Q}_W$ generated by the vertex $HWW$ emerged from an effective Lagrangian were calculated and estimated. A calculation of contributions to $WW\gamma$ from BSM scalar particles originated in the context of the Georgi-Machacek model~\cite{AHT} was carried out by the authors of Ref.~\cite{GeMa}.
\\

Our expressions for the electromagnetic form factors $\Delta\kappa$ and $\Delta Q$, given in Eqs.~(\ref{totDk})-(\ref{totDQ}), share the generic structure $\Delta f=\sum_\alpha\sum_{k=1}^6|{\cal B}_{\alpha k}|^2\Delta f^{\alpha k}$, so they can be written as
\begin{equation}
\Delta f=\sum_\alpha\sum_{k=1}^3\Big(|{\cal B}_{\alpha \nu_{k}}|^2\Delta f^{\alpha\nu_{k}}+|{\cal B}_{\alpha N_{k}}|^2\Delta f^{\alpha N_{k}}\Big),
\label{DfnuplusN}
\end{equation}
where the contributions from light and heavy neutrinos, $\nu_k$ and $N_k$, have been separated from each other. We express the $3\times3$ matrix $\xi$, defined in Eq.~(\ref{xidef}), as
\begin{equation}
\xi= \hat\rho X,
\label{theX}
\end{equation}
where $X$ is a $3\times3$ matrix whose largest entry has magnitude $1$, whereas $\hat\rho$ is some positive real number which equals the modulus of the largest entry of $\xi$, so $\hat\rho<1$. Thus, the $3\times3$ matrices ${\cal B}_\nu$ and ${\cal B}_N$, whose entries are ${\cal B}_{\alpha\nu}$ and ${\cal B}_{\alpha N}$, respectively, are expressed as
\begin{eqnarray}
&&
{\cal B}_\nu=V^\ell\textstyle\Big( {\bf 1}_3-\frac{1}{2}\hat\rho^2XX^\dag \Big),
\label{Bnuapproximation}
\\ \nonumber \\
&&
{\cal B}_N=V^\ell\,X\textstyle\Big( \hat\rho\cdot{\bf 1}_3-\frac{1}{2}\hat\rho^3X^\dag X \Big)J^*.
\label{BNapproximation}
\end{eqnarray}
The presence of the matrices $X$ and $J$ anticipates the involvement of a large number of parameters. We have tested different $X$ textures to estimate heavy-neutrino contributions to $\Delta\kappa$ and $\Delta Q$, but found no important variations of our results. So, for the sake of practicality, and aiming at an estimation of the new-physics effects carried by our analytical results, we take $X={\bf 1}_3$ and $J={\bf 1}_3$. As we commented before, in Section~\ref{theory}, the lepton-mixing matrix $V^\ell$, which appears in Eqs.~(\ref{Bnuapproximation}) and (\ref{BNapproximation}), needs not to be unitary. However, for practical purposes, in our estimations we assume that this matrix is approximately unitary, and then we use it as if it was the PMNS matrix, that is, we take $V^\ell\to{\cal U}_{\rm PMNS}$.
\\

The PMNS matrix is given, in any context of Majorana neutrinos, as ${\cal U}_{\rm PMNS}={\cal U}_{\rm D}\,{\cal U}_{\rm M}$. On the one hand, the $3\times3$ unitary matrix ${\cal U}_{\rm D}$ can be conveniently parametrized in terms of three mixing angles and one phase as~\cite{GiKi}
\begin{widetext}
\begin{equation}
{\cal U}_{\rm D}=
\left(
\begin{array}{ccc}
c_{12}c_{13} & s_{12}s_{13} & s_{13}e^{-i\delta_{\rm D}}
\\
-s_{12}c_{23}-c_{12}s_{23}s_{13}e^{i\delta_{\rm D}} & c_{12}c_{23}-s_{12}s_{23}s_{13}e^{i\delta_{\rm D}} & s_{23}c_{13}
\\
s_{12}s_{23}-c_{12}c_{23}s_{13}e^{i\delta_{\rm D}} & -c_{12}s_{23}-s_{12}c_{23}s_{13}e^{i\delta_{\rm D}} & c_{23}c_{13}
\end{array}
\right),
\end{equation}
\end{widetext}
where the standard compact notation $c_{jk}=\cos\theta_{jk}$, $s_{jk}=\sin\theta_{jk}$, for the cosine and the sine of the mixing angles $\theta_{12}$, $\theta_{23}$, and $\theta_{13}$ has been employed. Also, there lies the parameter $\delta_{\rm D}$, commonly referred to as the Dirac phase. On the other hand, the factor ${\cal U}_{\rm M}$ is the diagonal $3\times3$ matrix~\cite{GiKi}
\begin{equation}
{\cal U}_{\rm M}={\rm diag}(1,e^{i\phi_2},e^{i\phi_3}),
\end{equation}
with $\phi_2$ and $\phi_3$ the Majorana phases, only occurring as long as neutrinos correspond to Majorana fields. The Particle Data Group (PDG) recommends the following values for the neutrino-mixing angles~\cite{PDG}:
\begin{eqnarray}
&&\sin^2\theta_{12}=0.307\pm0.013,
\\
&&\sin^2\theta_{23}=0.546\pm0.0021,
\\
&&\sin^2\theta_{13}=0.0220\pm0.0007,
\end{eqnarray}
which we used for our numerical estimations. The value given by the PDG for $\theta_{12}$ is based on the 2016 measurement by the Super-Kamiokande Collaboration~\cite{SKtheta12}. Meanwhile, for the PDG value of $\theta_{23}$, data reported by the collaborations T2K~\cite{T2Ktheta23}, Minos+~\cite{MINOStheta23}, NOvA~\cite{NOvAtheta23}, IceCube~\cite{ICtheta23}, and Super-Kamiokande~\cite{SKtheta23} were considered. Furthermore, measurements given by Double Chooz~\cite{DCtheta13}, RENO~\cite{RENOtheta13,RENOtheta13o}, and Daya Bay~\cite{DBtheta13,DBtheta13o} were utilized by the PDG in order to set the afore-displayed value of $\theta_{13}$. In the case of the Dirac phase, we used 
\begin{equation}
\delta_{\rm D}=-\frac{\pi}{2}, 
\end{equation}
favored by the T2K Collaboration in their 2014 paper~\cite{T2KDirac}. A more recent analysis on the Dirac phase, by the same team, can be found in Ref.~\cite{T2KDiraco}. About the Majorana phases, we have taken the values $\phi_2=0$, $\phi_3=0$.
\\

Since the external photon has been assumed to be off shell, our expressions for the anomalies $\Delta\kappa$ and $\Delta Q$ are functions depending on the squared incoming momentum, $Q^2$, associated to the external photon line (see Fig.~\ref{cnvWWgamma}). Moreover, the contributions are also determined by the masses of the heavy neutrinos. In order to carry out an analysis, a heavy-neutrino mass spectrum must be defined. Such a spectrum is restricted to be nearly degenerate, for it to be compatible with loop-generated tiny light-neutrino masses, as required by the Majorana-neutrino model under consideration~\cite{Pilaftsis}. With this in mind, in what follows, our discussion considers a heavy-neutrino mass spectrum in which $m_{N_1}\approx m_{N_{\rm h}}$, $m_{N_2}\approx m_{N_{\rm h}}$, and $m_{N_3}\approx m_{N_{\rm h}}$, for some mass $m_{N_{\rm h}}$. While small variations among the masses of heavy neutrinos can be considered for the analysis, let us comment that our estimations showed practically no differences in the resulting contributions to $\Delta\kappa$ and $\Delta Q$.
\\

Hadron colliders bear the ability to probe the vertex $WW\gamma$. This is achieved through $W\gamma$ and $WW$ production resulting from proton-proton collisions, from which constraints on the form factors $\Delta\kappa$ and $\Delta Q$ can be established. In Refs.~\cite{ATLAS2017TGC,CMS2017TGC}, the CMS and the ATLAS Collaborations released upper bounds on $\Delta\kappa$ and $\Delta Q$, of order $10^{-2}$, from LHC data collected at a CME of $8\,{\rm TeV}$. An improved upper limit on $\Delta Q$, of order $10^{-3}$, was determined, just last year, from CMS-experiment data on $W\gamma$ production at a CME of $13\,{\rm TeV}$~\cite{CMS2021TGC}. The D0 Collaboration, at Tevatron, has also constrained these anomalies, reporting upper bounds of orders $10^{-1}$ on $\Delta\kappa$ and $10^{-2}$ on $\Delta Q$~\cite{D02012TGC}. Linear electron-positron colliders provide clean environments, suitable for high-precision studies, thus complementing hadron colliders in the identification of new-physics traces and their proper characterization. In particular, such devices are sensitive to the gauge coupling $WW\gamma$, as it participates in the process $e^+e^-\to W^+W^-$. The Large Electron-Positron Collider, better known as LEP and which is located at CERN, analyzed combined data, gathered from its four detectors, on $WW$ production from electron-positron collisions with a CME ranging from $130\,{\rm GeV}$ to $209\,{\rm GeV}$. Such an analysis yielded constraints of order $10^{-2}$ on both $\Delta\kappa$ and $\Delta Q$~\cite{LEP2013TGC}. While nowadays the LEP Collider is the largest of its kind, other similar devices are on the way. The International Linear Collider, commonly referred to by its acronym ILC, shall be able to establish constraints as restrictive as~\cite{LHCvsILC,ILCtechrepTGC,TGCatCEPC}
\begin{equation}
\left.
\begin{array}{r}
|\Delta\kappa|\leqslant6.1\times10^{-4},
\\
|\Delta Q|\leqslant8.4\times10^{-4},
\end{array}
\right\}
\text{ at a CME of 500 GeV},
\end{equation}
%%%%%%%%%%
\begin{equation}
\left.
\begin{array}{r}
|\Delta\kappa|\leqslant3.7\times10^{-4},
\\
|\Delta Q|\leqslant5.1\times10^{-4},
\end{array}
\right\}
\text{ at a CME of 800 GeV},
\end{equation}
on the anomalous triple gauge couplings, from $e^+e^-\to W^+W^-$. The authors of Ref.~\cite{BKGH} explored the processes $\gamma\gamma\to W^+W^-$, $e^+\gamma\to e^+\gamma^*\gamma\to e^+ W^-W^+$, and $e^+e^-\to e^+\gamma^*\gamma^*e^-\to e^+W^-W^+e^-$, then arriving at the conclusion that the CERN Compact Linear Collider, the CLIC, might reach upper bounds as stringent as $10^{-5}$ to $10^{-4}$ on $\Delta\kappa$ and $\Delta Q$. The Circular Electron-Positron Collider, also called the CEPC, is another electron-positron collider in plans. According to the investigation performed in Ref.~\cite{TGCatCEPC}, its sensitivity to trilinear gauge couplings shall allow it to set upper bounds of order $10^{-4}$ on $\Delta\kappa$ and $\Delta Q$.
\\

\begin{figure*}[ht]
\centering
\includegraphics[width=0.48\linewidth]{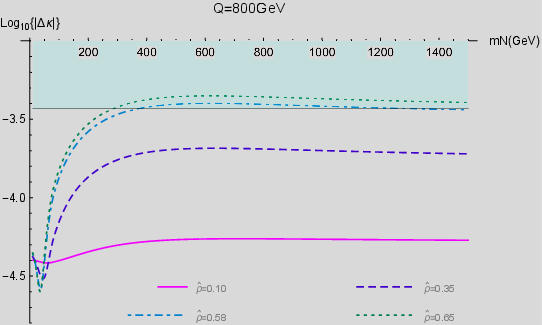}\hfil
\includegraphics[width=0.48\linewidth]{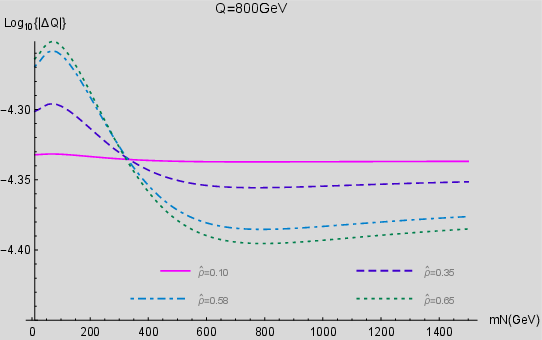}\par\medskip
\caption{\label{anomsvsmN} Contributions from neutrinos, both light and heavy, to the anomalies $\Delta\kappa$ (left graph) and $\Delta Q$ (right graph) as functions on the heavy-neutrino mass $m_{N_{\rm h}}$, in the range  $10\,{\rm GeV}\leqslant m_{N_{\rm h}}\leqslant1500\,{\rm GeV}$, at $\sqrt{Q^2}=800\,{\rm GeV}$, in logarithmic scale base-10, for different values of the parameter $\hat\rho$: $\hat\rho=0.10$ (solid); $\hat\rho=0.35$ (dashed); $\hat\rho=0.58$ (dot-dashed); and $\hat\rho=0.65$ (dotted).}
\end{figure*}

With the previous elements in mind, we refer the reader to the graphs of Fig.~\ref{anomsvsmN},
which display our estimations for the contributions from virtual neutrinos, both heavy and light, to the CP-even anomalies $\Delta\kappa$ and $\Delta Q$, in the framework set by the neutrino model discussed in Section~\ref{theory} of the present paper. Such contributions have been plotted as functions on the heavy-neutrino mass $m_{N_{\rm h}}$\footnote{The heavy-neutrino mass $m_{N_{\rm h}}$ has been denoted simpliy by $m_N$, in all the graphs of the present paper.}, within the range $10\,{\rm GeV}\leqslant m_{N_{\rm h}}\leqslant1500\,{\rm GeV}$, at $\sqrt{Q^2}=800\,{\rm GeV}$, as considered for analyses of the expected ILC sensitivity~\cite{LHCvsILC,ILCtechrepTGC,TGCatCEPC}. Aiming at a comparison between the orders of magnitude of the contributions, these graphs have been plotted in base-10 logarithmic scale. The Feynman diagrams contributing to $WW\gamma$ at one loop, which are displayed in Eq.~(\ref{vfsumofdiags}), involve vertices $W\nu_k\,l_\alpha$ and  $WN_k\,l_\alpha$, among which the former connects an external $W$ boson with the virtual loop fields $\nu_k$ and $l_\alpha$, where $m_W>m_\alpha+m_{\nu_k}$ is fulfilled. That this relation holds means that the resulting contributions from virtual light neutrinos to the on-shell form factors $\Delta\kappa(Q^2=0)$ and $\Delta Q(Q^2=0)$ are complex quantities. The rest of the diagrams, in which participating virtual neutrinos are heavy, produce on-shell form factors that are strictly real if heavy-neutrino masses meet the condition $m_W<m_\alpha+m_{N_k}$. However, the calculation has been performed with the external photon off the mass shell, so the contributions from virtual heavy neutrinos to the anomalies $\Delta\kappa$ and $\Delta Q$ are allowed to be complex quantities, even if virtual heavy-neutrino masses satisfy this condition. While these anomalies have real and imaginary parts, the contributions plotted in Fig.~\ref{anomsvsmN} correspond to the moduli $|\Delta\kappa|$ and $|\Delta Q|$. Each curve in the graphs of Fig.~\ref{anomsvsmN} corresponds to some value of the parameter $\hat\rho$, defined in Eq.~(\ref{theX}): solid curves arise for $\hat\rho=0.10$; the value $\hat\rho=0.35$ generates dashed plots; dot-dashed curves are associated to the value $\hat\rho=0.58$; and $\hat\rho=0.65$ yields dotted curves. A horizontal line, corresponding to the ILC upper bound $|\Delta\kappa|\leqslant3.7\times10^{-4}$ as estimated in Ref.~\cite{TGCatCEPC}, has been added to the left graph, which displays contributions $|\Delta\kappa|$. The upper shaded region, limited from below by this ILC-bound line, comprehend those $\Delta\kappa$ values which shall be accessible to the ILC. Conversely, the non-shaded lower region comprises the set of values out of the reach of the expected sensitivity for this collider. According to the left graph of Fig.~\ref{anomsvsmN}, couplings with $\hat\rho$ below $\sim0.5$ shall be beyond the reach of the ILC, at $\sqrt{Q^2}=800\,{\rm GeV}$. For instance, in this graph, the plots given by values $\hat\rho=0.58$ and $\hat\rho=0.65$ enter the ILC-sensitivity region for a range of masses $m_{N_{\rm h}}\gtrsim350\,{\rm GeV}$. This graph also shows that, even in the most optimistic context among those considered for the present paper, the BSM effects from the Majorana-neutrino model are smaller than the SM contribution by about one order of magnitude. No ILC-bound line has been included in the right graph of Fig.~\ref{anomsvsmN}, corresponding to the $|\Delta Q|$ anomaly, because the corresponding contributions are way below ILC expected sensitivity. At this point, it is worth mentioning Ref.~\cite{CMSmNvsB}, by the CMS Collaboration. In that work, the coefficients ${\cal B}_{eN_k}$ and ${\cal B}_{\mu N_k}$, for some heavy neutrino $N_k$, were investigated. This was carried out by searching for the decay of a Majorana heavy neutral lepton, with mass $1\,{\rm GeV}<m_{N_k}<1200\,{\rm GeV}$, into a SM charged lepton and the SM $W$ boson. For such a range of heavy-neutrino masses, upper bounds were established on $|{\cal B}_{eN_k}|^2$ and $|{\cal B}_{\mu N_k}|^2$, displayed in the planes $(m_{N_k},{\cal B}_{eN_k})$ and $(m_{N_k},{\cal B}_{\mu N_k})$. Recall that, according to Eq.~(\ref{BNapproximation}), the size of such quantities, which are entries of the $3\times3$ matrix ${\cal B}_N$, is given by the parameter $\hat\rho$. Keeping this in mind, we observe that the values $\hat\rho=0.58,\,0.65$ are compatible with the bounds on $|{\cal B}_{eN_k}|$, of Ref.~\cite{CMSmNvsB}, as long as $m_{N_k}\gtrsim850\,{\rm GeV}$. Regarding the bounds on $|{\cal B}_{\mu N_k}|$, the values $\hat\rho=0.58,\,0.65$ are allowed for $m_{N_k}\gtrsim1000\,{\rm GeV}$.
\\

\begin{figure*}[ht]
\centering
\includegraphics[width=0.48\linewidth]{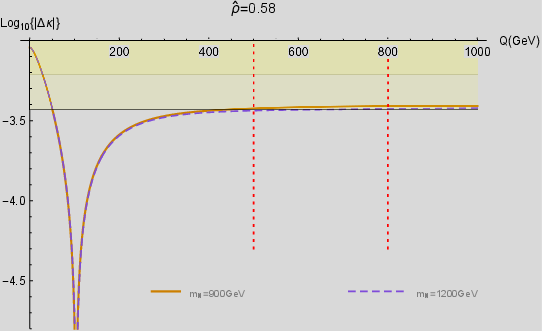}\hfil
\includegraphics[width=0.48\linewidth]{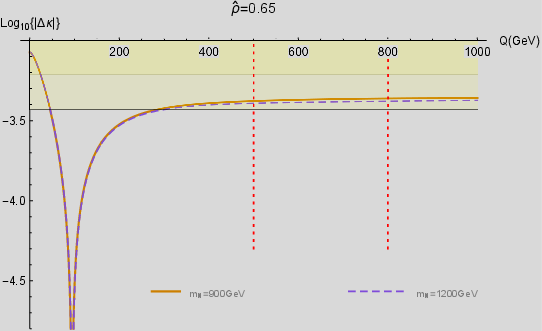}\par\medskip
\caption{\label{Dklambdas} $|\Delta\kappa|$ plotted against $\sqrt{Q^2}$, in the range $0\,{\rm GeV}\leqslant\sqrt{Q^2}\leqslant1000\,{\rm GeV}$, for the values $\hat\rho=0.58$ (left graph) and $\hat\rho=0.65$ (right graph). The curves have been plotted for heavy-neutrino masses $m_{N_{\rm h}}=900\,{\rm GeV}$ (solid curves), and $m_{N_{\rm h}}=1200\,{\rm GeV}$ (dashed curves). Dotted vertical lines represent the values $\sqrt{Q^2}=500\,{\rm GeV}$ and $\sqrt{Q^2}=800\,{\rm GeV}$, whereas solid horizontal lines stand for ILC sensitivity at $\sqrt{Q^2}=500\,{\rm GeV}$ (upper line) and $\sqrt{Q^2}=800\,{\rm GeV}$ (lower line).}
\end{figure*}

Our preceding discussion indicates that the plausibility of an ILC measurement of the anomaly $\Delta\kappa$, produced by the virtual Majorana neutrinos of the new-physics model under consideration, is more promising than the observation of a $\Delta Q$ effect. For this reason, from here on our discussion develops around the $\Delta\kappa$ coupling. In the framework of the present investigation, the participation of the parameter $\hat\rho$ is weighty in the definition of the size of the contributions, so it plays a significant role in whether the ILC shall be able to sense it or not. To further illustrate this, we show the graphs of Fig.~\ref{Dklambdas},
in which our estimations for the anomaly contribution $\Delta\kappa$ have been plotted for a couple of different values of $\hat\rho$, namely, $\hat\rho=0.58$ for the left graph and $\hat\rho=0.65$ in the case of the right graph. The plots of this figure, all of them given in base-10 logarithmic scale, are displayed as functions on the squared incoming momentum of the external photon $Q^2$, or more precisely in terms of $\sqrt{Q^2}$, for values running within $0\,{\rm GeV}\leqslant\sqrt{Q^2}\leqslant1000\,{\rm GeV}$. Both graphs include a couple of horizontal lines, which represent attainable ILC $\Delta\kappa$ sensitivity~\cite{LHCvsILC,ILCtechrepTGC,TGCatCEPC}, lying at $6.1\times10^{-4}$ for $\sqrt{Q^2}=500\,{\rm GeV}$ and $3.7\times10^{-4}$ for $\sqrt{Q^2}=800\,{\rm GeV}$. Moreover, two dotted vertical lines have been plotted in both graphs of the figure, corresponding to the values $\sqrt{Q^2}=500\,{\rm GeV}$ and $\sqrt{Q^2}=800\,{\rm GeV}$. Each graph comprises two curves, each of which is associated to one of two considered values of the heavy-neutrino mass $m_{N_{\rm h}}$: solid curves come from the choice $m_{N_{\rm h}}=900\,{\rm GeV}$; and the neutrino mass $m_{N_{\rm h}}=1200\,{\rm GeV}$ has been utilized to plot the dashed curves. Note, from the graphs of this figure, that disparities among contributions arisen from different heavy-neutrino masses $m_{N_{\rm h}}$ do not amount to orders of magnitude, so they are not fairly large. The left graph of Fig.~\ref{Dklambdas}, with its plots corresponding to $\hat\rho=0.58$, shows that the effects of the contributions barely lie within ILC reach at $800\,{\rm GeV}$, whereas sensitivity of this collider at $\sqrt{Q^2}=500\,{\rm GeV}$ is not achievable. If $\hat\rho=0.65$, the contributions associated to the two heavy-neutrino spectra under consideration clearly enter ILC expected-sensitivity region at $\sqrt{Q^2}=800\,{\rm GeV}$. Again, the Majorana-neutrino contribution at $\sqrt{Q^2}=500\,{\rm GeV}$ is clearly beyond ILC sensitivity.
\\

\begin{figure}[ht]
\centering
\includegraphics[width=8.5cm]{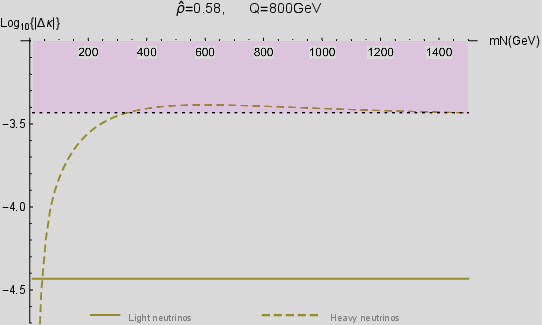}
\caption{\label{lightVSheavy} Plots, in logarithmic scale base-10, of light- and heavy-neutrino contributions $|\Delta\kappa_\nu|$ and $|\Delta\kappa_N|$, with respect to heavy-neutrino mass $m_{N_{\rm h}}$, within $10\,{\rm GeV}\leqslant m_{N_{\rm h}}\leqslant1500\,{\rm GeV}$, at $\sqrt{Q^2}=800\,{\rm GeV}$ and $\hat\rho=0.58$.}
\end{figure}

Up to this point, our discussion of the contributions from Feynman diagrams involving virtual Majorana neutrinos to the $WW\gamma$ vertex has been carried out by considering all the neutrinos, both light and heavy, at once. A comparison among the contributions from the virtual light neutrinos with those produced by the heavy ones is opportune, since it may be helpful in the discrimination and proper identification of their effects. We split our result for the CP-even anomalous coupling $\Delta\kappa$ into two terms as $\Delta\kappa=\Delta\kappa_\nu+\Delta\kappa_N$, with $\Delta\kappa_\nu$ the contribution generated by virtual light neutrinos and $\Delta\kappa_N$ the contribution corresponding to virtual heavy neutrinos. Now consider Fig.~\ref{lightVSheavy}, 
which shows the moduli of such contributions, $|\Delta\kappa_\nu|$ and $|\Delta\kappa_N|$, separately. This graph has been performed for $\sqrt{Q^2}=800\,{\rm GeV}$, fixed, and $\hat\rho=0.58$. As in previous cases, the plots are displayed in base-10 logarithmic scale. The horizontal dotted line, included in this graph, delimits the sensitivity region for ILC at $\sqrt{Q^2}=800\,{\rm GeV}$. The $\Delta\kappa$ contributions have been plotted against the heavy-neutrino mass $m_{N_{\rm h}}$, within the interval $10\,{\rm GeV}\leqslant m_{N_{\rm h}}\leqslant1500\,{\rm GeV}$. The contribution from the light neutrinos is represented in this figure by the horizontal solid line, at $|\Delta\kappa_\nu|=3.7\times10^{-5}$, or $\log_{10}\{ |\Delta\kappa_\nu| \}=-4.43$ in the logarithmic scale. The dashed curve, on the other hand, depicts the behavior of the anomaly contribution $|\Delta\kappa_N|$, associated to the heavy neutrinos. This graph shows that most values of $m_{N_{\rm h}}$ yield heavy-neutrino contributions which are larger than those corresponding to light neutrinos by about one order of magnitude. Therefore, a measurement of a BSM anomaly at $\sqrt{Q^2}=800\,{\rm GeV}$, in the assumed context of neutrino physics, by the ILC should not be linked to light neutrinos, but to heavy neutrinos with nearly-degenerate masses in the hundreds of GeVs.

\section{Summary and conclusions}
\label{sumandconc}
Throughout the present paper, we have addressed the calculation, estimation, and discussion of effects on the vertex $WW\gamma$ from physics beyond the Standard Model, produced by virtual Majorana neutrinos participating in one-loop Feynman diagrams. To this aim, the external Standard-Model $W$ bosons have been taken on shell and the external photon has been assumed to be off the mass shell. The framework for the execution of this investigation has been a new-physics neutrino model in which both Dirac-like and Majorana-like mass terms are assumed to emerge after a couple of stages of spontaneous symmetry breaking occurring in the context of some high-energy description of nature. While conditions are set for a type-1 seesaw mechanism to naturally take place, we rather follow Ref.~\cite{Pilaftsis}, in which this mechanism of neutrino-mass generation is avoided. Current upper bounds on the masses of light neutrinos are as stringent as $m_{\nu_k}\lesssim0.8\,{\rm eV}$~\cite{KATRINnumass}, so light-neutrino masses defined at tree level by a seesaw mechanism push the masses of heavy neutrinos into very large values. By imposing a condition of light-neutrino mass cancellation at tree level, the Majorana-neutrino model under consideration defines such masses radiatively, which allows for smaller masses of heavy neutrinos~\cite{Pilaftsis}. In this context, heavy-neutrino mass spectra are constrained to be nearly degenerate, in order to get tiny masses for light neutrinos~\cite{Pilaftsis}.
\\

Our calculation engendered contributions to the Lorentz-covariant parametrization of the gauge vertex $WW\gamma$~\cite{HPZH,BaZe}, which includes, in general, both CP-even and CP-odd effects, and which abides by electromagnetic gauge invariance. A CP-odd contribution $\Delta\tilde{Q}$ seemed to appear at first glance, but usage of Schouten identities yielded its exact cancellation, thus leading to the conclusion that the contributions to $WW\gamma$ from this model lack CP-odd effects. Contributions to CP-even anomalies $\Delta\kappa$ and $\Delta Q$ were found to be, on the other hand, nonzero and free of ultraviolet divergences. By virtue of the off-shell assumption on the external photon, our calculation gave rise to anomalous couplings $\Delta\kappa$ and $\Delta Q$ depending on the external-photon squared momentum $Q^2$, though notice that such quantities are determined by heavy-neutrino masses as well. Furthermore, the resulting anomaly contributions turned out to be complex quantities for several $Q^2$ values.
\\

We carried out numerical estimations that led us to conclude that most values of the heavy-neutrino masses yield effects which dominate over those coming from the virtual light neutrinos, with a disparity amounting to $\sim1$ order of magnitude. Our discussion comprehended the analysis of both sorts of virtual-neutrino contributions, which were estimated and discussed in the light of the sensitivity of the International Linear Collider, expected to set bounds of order ${\cal O}(10^{-4})$ on these anomalous couplings~\cite{LHCvsILC,ILCtechrepTGC,TGCatCEPC} through $WW$ production from the process $e^+e^-\to W^+W^-$. A parameter $\hat\rho>0$ was defined to characterize the size of the couplings $WN_kl_\alpha$, among the Standar-Model $W$ boson, the heavy neutrinos and the Standar-Model charged leptons, with the constraint $\hat\rho<1$. The $\hat\rho$ parameter turned out to be quite relevant in the determination of the size of the contributions. We found that $\hat\rho$ values larger than $\sim0.5$ give rise to contributions to the anomaly $\Delta\kappa$ as large as $\sim10^{-3}$, which are smaller than the Standard-Model contribution by $\sim1$ order of magnitude. Moreover, an anomaly contribution this size would be attainable by the International Linear Collider, though $e^+e^-\to W^+W^-$ at a center-of-mass energy of $800\,{\rm GeV}$. According to our analysis, the observation of such a new-physics effect should be associated to heavy neutrinos with masses in the range of the hundreds of GeVs. The CP-even anomaly $\Delta Q$, on the other hand, receives contributions as large as $\lesssim10^{-4}$, which are too small to be observed by the International Linear Collider.
\\\\\\\\

\section*{Acknowledgements}
\noindent
We acknowledge financial support from Conacyt (M\'exico).

\appendix*

\section{Explicit expressions of the form-factor contributions}
In this Appendix we provide the explicit expressions of the form-factor contributions $\Delta\kappa^{\alpha k}$ and $\Delta Q^{\alpha k}$, produced by the Feynman diagram involving the $\alpha$-th virtual charged lepton and the $k$-th virtual neutrino, in accordance with Eqs.~(\ref{totDk}) and (\ref{totDQ}). Since the calculation was effectuated by means of the Passarino-Veltman tensor-reduction method~\cite{PaVe}, these contributions are given in terms of scalar functions~\cite{tHVescalar}, which we denote as
\begin{eqnarray}
&&
A_0^{(1)}=A_0(m_\alpha^2),
\\ &&
A_0^{(2)}=A_0(m_k^2),
\\ &&
B_0^{(1)}=B_0(m_W^2,m_\alpha^2,m_k^2),
\\ &&
B_0^{(2)}=B_0(Q^2,m_\alpha^2,m_\alpha^2),
\\ &&
C_0=C_0(m_W^2,m_W^2,Q^2,m_\alpha^2,m_k^2,m_\alpha^2).
\end{eqnarray}
The form factors have the structure
\begin{eqnarray}
&&
\Delta\kappa^{\alpha k}=\frac{g^2}{(4\pi)^2}\Big(\xi^{(1)}_{\alpha k}A_0^{(1)}+\xi^{(2)}_{\alpha k}A_0^{(2)}
\nonumber\\ &&\hspace{1cm}
+\eta^{(1)}_{\alpha k}B_0^{(1)}+\eta^{(2)}_{\alpha k}B_0^{(2)}+\beta_{\alpha k}C_0\Big),
\\\nonumber\\ &&
\Delta Q^{\alpha k}=\frac{g^2}{4\pi^2}\Big(\zeta^{(1)}_{\alpha k}A_0^{(1)}+\zeta^{(2)}_{\alpha k}A_0^{(2)}
\nonumber\\ &&\hspace{1cm}
+\lambda^{(1)}_{\alpha k}B_0^{(1)}+\lambda^{(2)}_{\alpha k}B_0^{(2)}+\omega_{\alpha k}C_0\Big),
\end{eqnarray}
with the definitions
\begin{widetext}
%%%%%%%%%%
\begin{eqnarray}
\xi^{(1)}_{\alpha k}=\frac{1}{(D-1) m_W^2 \left(Q^2-4m_W^2\right){}^2}\Big(4 m_W^2 \big((D+2)m_k^2
\nonumber \\
-(D+2) m_{\alpha }^2+Dm_W^2\big)+Q^2\left(-m_k^2+m_{\alpha}^2+m_W^2\right)\Big),
\end{eqnarray}
%%%%%%%%%%
\begin{eqnarray}
\xi^{(2)}_{\alpha k}=\frac{1}{(D-1) m_W^2 \left(Q^2-4m_W^2\right){}^2}\Big(Q^2 \left((3-2 D)m_W^2+m_k^2-m_{\alpha }^2\right)
\nonumber \\
+4m_W^2 \left(-(D+2) m_k^2+(D+2)m_{\alpha }^2+(D-4)m_W^2\right)\Big),
\end{eqnarray}
%%%%%%%%%%
\begin{eqnarray}
\eta^{(1)}_{\alpha k}=\frac{1}{(D-2) (D-1) m_W^2 \left(4m_W^2-Q^2\right){}^3}\Big((D-2) Q^4 \big(2 m_{\alpha}^2 \left((D-2)m_W^2-m_k^2\right)+2 (D-2) m_k^2m_W^2+m_k^4
\nonumber \\
+m_{\alpha}^4+m_W^4\big)+4 Q^2 m_W^2\big(2 m_{\alpha }^2 \left((3D-7) m_k^2+((11-3 D) D-12)m_W^2\right)-2 ((D-9) D+12) m_k^2m_W^2
\nonumber \\
+(7-3 D) m_k^4+(7-3 D)m_{\alpha }^4+(D (2 D-7)+9)m_W^4\big)+16 m_W^4\big(2m_{\alpha }^2 \big(((D-7) D+9)m_W^2
\nonumber \\
-(D (D+2)-6) m_k^2\big)+2((D-5) D+7) m_k^2 m_W^2+(D(D+2)-6) m_k^4(D(D+2)-6)m_{\alpha }^4
\nonumber \\
-(D-4) (D-2)m_W^4\big)\Big),
\end{eqnarray}
%%%%%%%%%%
\begin{eqnarray}
\eta^{(2)}_{\alpha k}=\frac{-2}{(D-2) (D-1) \left(4m_W^2-Q^2\right){}^3}\Big(m_{\alpha }^2 \big(-4m_W^2\left(8 (D-1) m_k^2+(D (3D-10)+11) Q^2\right)
\nonumber \\
+(D-1) Q^2\left((D-3) Q^2-4 (D-1)m_k^2\right)+16 (D-3)^2m_W^4\big)+2 m_W^4 \big(8 (D-3)(D-1) m_k^2
\nonumber \\
+(D(D+2)-11)Q^2\big)+(D-1)^2Q^2m_k^2\left(2m_k^2+Q^2\right)+m_W^2\big(-4 (D-5) (D-1) Q^2 m_k^2
\nonumber \\
+16(D-1) m_k^4-(D-5) Q^4\big)+2(D-1) m_{\alpha }^4 \left((D-1)Q^2+8 m_W^2\right)\Big),
\end{eqnarray}
%%%%%%%%%%
\begin{eqnarray}
\beta_{\alpha k}=\frac{1}{(D-2) \left(4m_W^2-Q^2\right){}^3}\Big(-(D-2) Q^6 m_k^2+2 Q^4\big(m_k^2 \left(2 (D-1)m_{\alpha }^2+3 (D-4)m_W^2\right)-2 (D-1) m_k^4
\nonumber \\
+(D-2)m_W^2 \left(m_{\alpha }-m_W\right)\left(m_{\alpha}+m_W\right)\big)+4 Q^2\big(-(D-1) m_{\alpha }^4 \left(3m_k^2+5 m_W^2\right)+m_{\alpha }^2\big(2 (D+5) m_k^2 m_W^2
\nonumber \\
+3 (D-1)m_k^4+(3 D+1) m_W^4\big)+3 (D-5)m_k^4 m_W^2+(19-7 D) m_k^2m_W^4-(D-1) m_k^6+(D-1) m_{\alpha}^6
\nonumber \\
+(D-5) m_W^6\big)+32 m_W^2\left(m_{\alpha }-m_k\right)\left(m_k+m_{\alpha }\right)\big(m_k^2 \left((D-4) m_W^2-2m_{\alpha}^2\right)
\nonumber \\
+\left(m_{\alpha}-m_W\right) \left(m_{\alpha}+m_W\right) \left((D-3)m_W^2+m_{\alpha}^2\right)+m_k^4\big)\Big),
\end{eqnarray}
%%%%%%%%%%
\begin{eqnarray}
\zeta^{(1)}_{\alpha k}=\frac{-1}{(D-1) Q^2 \left(Q^2-4 m_W^2\right){}^2}\Big( m_k^2 \left(2 (D-1)m_W^2+Q^2\right)
\nonumber\\
-m_{\alpha }^2 \left(2 (D-1) m_W^2+Q^2\right)+2 m_W^2 \left((D-3)m_W^2+Q^2\right) \Big),
\end{eqnarray}
%%%%%%%%%%
\begin{eqnarray}
\zeta^{(2)}_{\alpha k}=\frac{1}{(D-1) Q^2 \left(Q^2-4 m_W^2\right){}^2}\Big( m_k^2 \left(2 (D-1)m_W^2+Q^2\right)
\nonumber\\
-m_{\alpha }^2 \left(2 (D-1) m_W^2+Q^2\right)+D Q^2 m_W^2-2 (D-1) m_W^4 \Big),
\end{eqnarray}
%%%%%%%%%%
\begin{eqnarray}
\lambda^{(1)}_{\alpha k}=\frac{1}{(D-2) (D-1) Q^2 \left(Q^2-4m_W^2\right){}^3}\Big( 2 m_k^2 \big((D-2)(Q^2)^2 m_{\alpha }^2+Q^2 m_W^2 \left(4 (4-3 D) m_{\alpha }^2+(2 D-3) Q^2\right)
\nonumber\\
+2 m_W^4 \left(((D-4)D+5) Q^2-4 (D-3) (D-1) m_{\alpha }^2\right)\big)+m_k^4 \big(4 (3 D-4) Q^2 m_W^2+8 (D-3) (D-1)m_W^4
\nonumber\\
-(D-2) (Q^2)^2\big)+\left(m_W^2-m_{\alpha }^2\right) \big((D-2) (Q^2)^2 m_{\alpha }^2+4 m_W^4\left(((D-3) D+1) Q^2-2 (D-3) (D-1) m_{\alpha }^2\right)
\nonumber\\
+Q^2 m_W^2 \left(4 (4-3 D) m_{\alpha }^2+DQ^2\right)-8 (D-3) (D-1) m_W^6\big) \Big),
\end{eqnarray}
%%%%%%%%%%
\begin{eqnarray}
\lambda^{(2)}_{\alpha k}=\frac{1}{2(D-2) (D-1) Q^2\left(Q^2-4 m_W^2\right){}^3}\Big( m_W^2 \big(4 m_{\alpha }^2\left(2 \left(D^2+D-10\right) Q^2 m_W^2-8 (D-3) m_W^4+3 (Q^2)^2\right)
\nonumber\\
-4 (D-1) m_k^4 \left((D+2) Q^2-4m_W^2\right)-4 (D-1) m_k^2 \big(-2 (D+2) Q^2 m_{\alpha }^2+2 m_W^2 \left((D-6) Q^2+4 m_{\alpha}^2\right)
\nonumber\\
+8 m_W^4+3 (Q^2)^2\big)-8 (D-3) (Q^2)^2 m_W^2-4 (D-1) m_{\alpha }^4 \left((D+2) Q^2-4m_W^2\right)-4 ((D-3) D+6) Q^2 m_W^4
\nonumber\\
+16 (D-1) m_W^6+(D-4) (Q^2)^3\big) \Big),
\end{eqnarray}
%%%%%%%%%%
\begin{eqnarray}
\omega_{\alpha k}=-\frac{1}{(D-2) Q^2\left(Q^2-4 m_W^2\right){}^3}\Big( m_W^2 \big(2 m_k^6\left((D+2) Q^2-4 m_W^2\right)+m_k^4 \big(-6 (D+2) Q^2 m_{\alpha }^2
\nonumber\\
+2 m_W^2 \left((D-16) Q^2+12m_{\alpha }^2\right)+(D+8) (Q^2)^2+24 m_W^4\big)+2 m_k^2 \big((D-6) (Q^2)^2 m_W^2
\nonumber\\
+3 m_{\alpha }^4\left((D+2) Q^2-4 m_W^2\right)-(D-14) Q^2 m_W^4-m_{\alpha }^2 \left(2 (D-6) Q^2 m_W^2+(D+4) (Q^2)^2+8m_W^4\right)
\nonumber\\
-12 m_W^6+(Q^2)^3\big)+\left(m_W^2-m_{\alpha }^2\right){}^2 \left(-2 (D+2) Q^2 m_{\alpha}^2+m_W^2 \left(8 m_{\alpha }^2-2 D Q^2\right)+D (Q^2)^2+8 m_W^4\right)\big) \big).
\end{eqnarray}
%%%%%%%%%%
\end{widetext}

\end{document}